**Nonrelativistic and Nonstationary Effective Mass Bound-Spectra Analysis of Squared Class Trigonometric Potentials Through the Point Canonical Formalism**

Metin Aktaş[(*)]

Ankara Yildirim Beyazit University, School of Engineering and Natural Sciences, Department of Energy Systems Engineering, 06010, Ankara, Turkey

**Abstract**

The present paper engages in a particular attempt to acquire exact analytical eigensolutions of the position-dependent effective mass (PDEM) Schrödinger equation for a variety of squared style trigonometric potentials. The algebraic process entitled as the point canonical transformation (PCT) approach is implemented in the course of study. Certain spatially varying effective mass configurations have been utilized in establishing of the target system. Then, performing the systematic computational procedures enables us to determine not only the possible explicit forms of both discrete energy spectra and their corresponding wavefunctions but also canonical counterparts of original potentials, involved in the framework of PDEM based quantum systems.

**Introduction**

In the past few decades of the literature, considerable attempts have been employed to achieve the solutions of Schrödinger, Klein-Gordon and Dirac equations involving non-stationary mass dependence formalism due to providing scientific viewpoints in analysis of the various mechanisms concerning particularly small scale physics. Besides ensuring the quantum mechanical description of some physical structures at micro and nano scales, position-dependent effective mass (PDEM) formulation of Schrödinger equation which is also widely used for quantum semiconductor heterostructures [1-7], quantum dots and liquids [8], $^3He$ and $^4He$ structures [9] has been acquired. For instance, this concept is applied especially in the determination of electronic transport properties of semiconductors, the introduction of the effective interaction pictures of nuclear particles [10] as well as the dynamical properties of neutron superfluids in the neutron stars [11].

It is pointed out that the exact solvability condition involved in both nonrelativistic and relativistic frameworks of quantum mechanical systems is very crucial factor as it can serve as a useful model to describe well these formations. Despite limited to achieving fully solvable quantum structures, the explicit solution of the Schrödinger equation with some spatially dependent mass distributions can be obtained. A great many alternative mathematical methods have been proposed in the literature to get exact and analytical solutions for equations having non-constant mass forms. For instance, these are point canonical transform, Lie algebraic and group theoretical approaches as well as alternative perspectives [12-39], parity-time (PT)-symmetric [40-44] and supersymmetric quantum mechanical (SUSYQM) formalisms etc [45-47]. In addition to the quantum mechanical problem of a variable mass particle under the concept of instantaneous Galilean invariance, the path integral approach [48, 49] and some other attempts to recognize various physical attributes [50-60] are also presented. Recently, several applications have been accomplished in the solution of Schrödinger [61-71], Klein-Gordon and Dirac equations etc [72-90], which are defined by changing masses.

[(*)]E-mail address: metinaktas01@gmail.com

Figuring out the canonical map solution of Schrödinger wave equation (SE) by proposing some novel parametric and rational form positionally variable mass configurations like $f(z) = P(z)/Q(z)$ is one of the major motivations of this research paper. The original map and the associated with the target equivalent one (counterpart) defined by the mass distribution function is provided by performing the canonical map process. To investigate the target spectrum, the point canonical transformation procedure is implemented algebraically. It is important here to note that the final configuration includes not only the energy spectra of the target system, but also the corresponding wavefunctions in terms of reference (source) potential. In the study of this research article, exact solutions of SE characterized by non-stationary effective masses will be handled by the PCT technique. The squared type trigonometric potentials such as *tangent*, *cotangent* [91, 92] and *trigonometric Pöschl-Teller* ones [93] will be examined respectively. By selecting several new form mass distribution functions depending on the spatial coordinate like $m = m(z)$ [94], then explicit and canonical maps of the target potentials as well as the eigenfunctions are generated analytically.

The outline of the article will be organized as follows: First, general description of the PCT method is presented by mapping phase of the spatially dependent effective mass Schrödinger equation. The third, fourth and fifth sections cover mainly the straightforward PCT calculations for the wave equation involving in various novel forms of mass distribution functions with the parameters especially $\alpha, \beta, \gamma$ and $\delta$ as well. The final section includes analysis and discussion of the results finding in the study.

1. **PCT Formulation of the Spatially Variable Effective Mass Schrödinger Equation**

The first step is the formulation of the spatially-dependent effective mass Schrödinger equation with the help of this well-known procedure. In order to do this, we first consider the Schrödinger's Hamiltonian in the case of position-dependent mass as

$$-\frac{1}{2}\left[\nabla_z \left(\frac{1}{M(z)}\right) \nabla_z \right]\Psi(z) + [E - U(z)]\Psi(z) = 0, \qquad (1)$$

where $M(z) = m_0\, m(z)$, and constant $m_0$. Then, its explicit form will be

$$\frac{d^2\Psi}{dz^2} - \left(\frac{m'}{m}\right)\frac{d\Psi}{dz} + 2m[E - U(z)]\Psi(z) = 0. \qquad (2)$$

In the equation, the atomic units ($\hbar = c = 1$) is chosen. Also, prime and double prime factors in the wavefunction for its first and second derivatives with respect to $z$. On the other hand, the one dimensional SE concerning the constant mass is written

$$\frac{d^2\Phi}{dy^2} + 2[\mathcal{E} - U(y)]\Phi(y) = 0. \qquad (3)$$

By making the transformation $y \to z$ through a mapping function $y = f(z)$, and the wavefunction is regarded as

$$\Phi(y) = \rho(z)\,\Phi(z), \tag{4}$$

then the stationary mass SE can be transformed into the new form

$$\frac{d^2\Phi}{dz^2} + 2\left(\frac{\rho'}{\rho} - \frac{f''}{2f'}\right)\frac{d\Phi}{dz}$$

$$+ \left\{\left(\frac{\rho''}{\rho} - \frac{f''}{f'}\frac{\rho'}{\rho}\right) + 2(f')^2[\mathcal{E} - U(f(z))]\right\}\Phi(z) = 0. \tag{5}$$

When comparing the left-hand sides of equations (2) and (5) term-by-term, then one can get the following conditions

$$\rho(z) = \sqrt{\frac{f'(z)}{m(z)}}, \tag{6}$$

and

$$[E - U(z)] = \frac{(f')^2}{m}[\mathcal{E} - U(f(z))] + \frac{1}{2m}\left(\frac{\rho''}{\rho} - \frac{f''}{f'}\frac{\rho'}{\rho}\right) \tag{7}$$

By setting $m = (f')^2$ and substituting it in the final equation, it will bring about the target problem including the energy spectra of the bound states, potential and wavefunction respectively

$$E_n = \mathcal{E}_n,$$

$$U(z) = U(f(z)) + \frac{1}{8m}\left[\frac{m''}{m} - \frac{7}{4}\left(\frac{m'}{m}\right)^2\right], \tag{8}$$

$$\Psi_n(z) = \sqrt[4]{m(z)}\,\Phi_n(f(z)).$$

This algebraic approach can be applied in some cases to determine their target *equivalents* and wavefunctions explicitly. It is also pointed out here that applying the well-known strategic process preserves the structure of the wavefunction in the target system having the same class as that of the reference problem.

In the following steps, some spatial mass distribution functions like $m = m(z)$ and the reference (source) potentials including squared *tangent*, *cotangent* [91, 92] as well as *trigonometric Pöschl-Teller* ones [93] are going to be considered respectively. In addition, their target eigenspectra configurations will be generated explicitly by applying the PCT method given in this section.

## 2. PCT Calculations For Mass Distribution – I

Let us first consider the following mass distribution form given [94]

$$m = m_1(z) = \left[\frac{\delta}{(\alpha+\beta z+\gamma z^2)}\right]^2, \tag{9}$$

where $\alpha, \beta, \gamma$ and $\delta$ are mass parameters. In this case, its mapping function becomes

$$y = \bar{f}(z) = \int \sqrt{m_1}\, dz$$

$$= \frac{2\delta}{\sqrt{\Delta}} \arctan\left(\frac{\beta+2\gamma z}{\sqrt{\Delta}}\right), \tag{10}$$

with the function $z = \frac{1}{2\gamma}\left[-\beta + \sqrt{\Delta} \tan\left(\frac{\sqrt{\Delta}}{2\delta}y\right)\right]$, where $\Delta = 4\alpha\gamma - \beta^2$ indicates the discriminant satisfying the condition $\Delta > 0$.

## 3. Squared Tangent Potential (STP) Case

First, we will pay our attention to the squared tangent potential (STP) case

$$U(y) = U_0 \tan^2 y, \tag{11}$$

where $U_0 = \mu(\mu - 1)$ is the potential coefficient including the parameter $\mu \geq 1$ and $y \in \left(-\frac{\pi}{2}, \frac{\pi}{2}\right)$. This is the original (source) potential with the energy spectra corresponding to the wavefunctions [91].

$$E_{2n}(\mu) = [4n(n+\mu) + \mu], \qquad n = 0, 1, 2, \ldots$$

$$\Phi_{2n}(y; \mu) = C_n(\mu)(\cos^\mu y)\, {}_2F_1(-n,\, n+\mu;\, \mu+1/2;\, \cos^2 y) \tag{12}$$

They are the symmetric eigenvalues and eigenfunctions regarding for even parity solutions. In the equation, $C_n(\mu)$ is the normalizable coefficient, and ${}_2F_1$ stands for the Gauss hypergeometric function. In addition to this, anti-symmetric eigenspectrum and the corresponding eigenfunctions results for odd parity solutions are given [92]

$$E_{2n+1}(\mu) = [(2n+1)(2n+2\mu+1) + \mu], \qquad n = 0, 1, 2, \ldots$$

$$\Phi_{2n+1}(y; \mu) = B_n(\mu)(\sin y)(\cos^\mu y)$$

$$\times \,_2F_1(-n,\ n+\mu+1;\ \mu+1/2;\ \cos^2 y). \tag{13}$$

By plugging the mass distribution (9) and the mapping function (10) into the equation (8), the constructed target system for the even and odd parity solutions of the STP case takes the form respectively

$$U(z) = \bar{U}_0\,(\beta + 2\gamma z)^2$$

$$+ \frac{1}{32(\alpha+\beta z+\gamma z^2)^4\,\delta^2}\left[-(7\beta + 8\gamma + 2\gamma z) + \frac{8(\beta+2\gamma z)^2}{(\alpha+\beta z+\gamma z^2)}\right]$$

$$\Psi_{2n}(z) = \sqrt{\frac{\delta}{(\alpha+\beta z+\gamma z^2)}}\;\Phi_{2n}\!\left(\bar{f}(z)\right), \tag{14}$$

and

$$E_{2n+1}(\mu) = \mathcal{E}_{2n+1}(\tau)$$

$$\Psi_{2n+1}(z) = \sqrt{\frac{\delta}{(\alpha+\beta z+\gamma z^2)}}\;\Phi_{2n+1}\!\left(\bar{f}(z)\right), \tag{15}$$

with $\bar{U}_0 = (\varkappa\, U_0)$ and $\varkappa = \left(\frac{4\delta^2}{\Delta}\right)$.

4. **Squared Cotangent Potential (SCP) Case**

It can be defined by

$$U(y) = U_0\,\cotan^2 y = \frac{U_0}{\tan^2 y}, \tag{16}$$

with the condition $y \in (0,\ \pi)$.

Similarly, this reference potential has also two identical solutions for both even and odd index parities, as in the case of STP. Following the similar algebraic method in solving of the SE for SCP, then they are written [91, 92]

$$\bar{\Phi}_{2n}(y + \pi/2\,;\mu) = (\sin^\mu y)\,_2F_1(-n,\ n+\mu;\ \mu+1/2;\ \sin^2 y) \tag{17}$$

eigenfunctions for the symmetric states and

$$\bar{\Phi}_{2n+1} = \Gamma_n \, \bar{\Phi}_{2n}(y + \pi/2; \mu = 3) = \Gamma_n \sum_{k=0}^{n} \frac{(-n)_k \, (n+3)_k}{k! \, (7/2)_k} \, (sin^{2k+3} y) \tag{18}$$

the anti-symmetric eigenfunctions and the normalizable coefficient $\Gamma_n$, n = 0, 1, 2, ... It should be mentioned here that the energy eigenvalues, but not eigenfunctions, of the Schrödinger's Hamiltonian for the STP on the symmetric interval $(-\pi/2, \pi/2)$ have precisely uniform as those of the Hamiltonian for the SCP case on the asymmetric interval $(0, \pi)$ [92].

By following the same procedure as in the above section, the even and odd parity target system solutions for (16) – (18) are obtained.

$$\bar{U}(z) = U(z) = \bar{U}_0 \, (\beta + 2\gamma z)^{-2}$$

$$+ \frac{1}{32(\alpha+\beta z+\gamma z^2)^4 \, \delta^2} \left[ -(7\beta + 8\gamma + 2\gamma z) + \frac{8 \, (\beta+2\gamma z)^2}{(\alpha+\beta z+\gamma z^2)} \right]$$

$$\bar{\Psi}_{2n}(z) = \sqrt{\frac{\delta}{(\alpha+\beta z+\gamma z^2)}} \, \bar{\Phi}_{2n}\left(\bar{f}(z)\right), \tag{19}$$

and

$$\bar{E}_{2n+1} = \bar{\mathcal{E}}_{2n+1}$$

$$\bar{\Psi}_{2n+1}(z) = \sqrt{\frac{\delta}{(\alpha+\beta z+\gamma z^2)}} \, \bar{\Phi}_{2n+1}\left(\bar{f}(z)\right). \tag{20}$$

It is pointed out here that the spectrum of this system and that of its original counterpart are identical.

## 5. Trigonometric Pöschl – Teller Potential (PTP) Case

Now, the general form of the trigonometric Pöschl-Teller (PTP) potential case is taken into account

$$U_{PT}(y) = \left( \frac{U_{01}}{sin^2 y} + \frac{U_{02}}{cos^2 y} \right)$$

$$= U_{01} \, cosec^2 \, y + U_{02} \, sec^2 \, y. \tag{21}$$

Here, coefficients of this potential are described as $U_{01} = U_0 \, \chi(\chi - 1)/2$ and $U_{02} = U_0 \, \lambda(\lambda - 1)/2$; $\chi \geq 1$, $\lambda \geq 1$ [93]. It is bounded in the region between the values of 0 and $\pi/2$. Solving the Schrödinger equation for (21) gives the energy eigenvalues as

$$\tilde{E}_n(\chi, \lambda) = \left[ \sqrt{\frac{U_0}{2}} \, (2n + \chi + \lambda) \right]^2, \tag{22}$$

and the wavefunctions

$$\widetilde{\Phi}_n(y; \chi, \lambda) = C_n(\sin^\chi y)(\cos^\lambda y) \, _2F_1(-n, \ n + \chi + \lambda, \ \chi + 1/2; \sin^2 y). \qquad (23)$$

Here $C_n$ is the normalization constant. Hence, energy spectra, target potential and wave functions can be developed by following equations (9) and (10)

$$\widetilde{U}(z) = \overline{U}_\omega \left[ \frac{U_{01}}{(\beta+2\gamma z)^2} + U_{02}(\beta + 2\gamma z)^2 \right]$$

$$+ \frac{1}{32(\alpha+\beta z+\gamma z^2)^4 \, \delta^2} \left[ -(7\beta + 8\gamma + 2\gamma z) + \frac{8\,(\beta+2\gamma z)^2}{(\alpha+\beta z+\gamma z^2)} \right]$$

$$\widetilde{\Psi}_n(z) = \sqrt{\frac{\delta}{(\alpha+\beta z+\gamma z^2)}} \ \widetilde{\Phi}_n\left(\overline{f}(z)\right), \qquad (24)$$

where the parameter is $\overline{U}_\omega = \left(1 + \frac{4\,\delta^2}{\Delta} U_0\right)$.

### 6. PCT Calculations For Mass Distribution – II

In this context, we consider the second kind mass distribution function [94]

$$m = m_2(z) = \left(\frac{\alpha z}{\beta^4 + \gamma^4 z^4}\right)^2. \qquad (25)$$

Thus it can be mapped by

$$y = \tilde{f}(z) = \int \sqrt{m_2} \, dz$$

$$= \frac{\alpha}{2\beta^2 \gamma^2} \, arctan\left(\frac{z^2}{\beta^2 \gamma^2}\right), \qquad (26)$$

giving the function $z = \beta \, \gamma \left[tan\left(\frac{2\beta^2 \gamma^2}{\alpha} y\right)\right]^{1/2}$.

### 7. STP Case

The mapping function yields the expressions

$$U(z) = \tilde{U}_0 \, z^4 - \left[\left(\frac{21}{8\,\alpha^2}\right)z^4 + \frac{5/(32\,\sigma)}{z^4} + \frac{5/(32\,\sigma)}{\beta^4\,\gamma^4}\right]$$

$$\Psi_{2n}(z) = \sqrt{\left(\frac{\alpha\,z}{\beta^4 + \gamma^4\,z^4}\right)} \, \Phi_{2n}\left(\tilde{f}(z)\right) \qquad (27)$$

$$\Psi_{2n+1}(z) = \sqrt{\left(\frac{\alpha\,z}{\beta^4 + \gamma^4\,z^4}\right)} \, \Phi_{2n+1}\left(\tilde{f}(z)\right),$$

with the coefficient of the potential $\tilde{U}_0 = (\sigma\,U_0)$ and with the same eigenspectrum and the potential parameter $\sigma = \left(\frac{\alpha^2}{4\,\beta^8\,\gamma^8}\right)$.

## 8. SCP Case

With the help of PCT procedure, the target system can be determined

$$U(z) = \tilde{U}_0 \, z^{-4} - \left[\left(\frac{21}{8\,\alpha^2}\right)z^4 + \frac{5/(32\,\sigma)}{z^4} + \frac{5/(32\,\sigma)}{\beta^4\,\gamma^4}\right]$$

$$\bar{\Psi}_{2n}(z) = \sqrt{\left(\frac{\alpha\,z}{\beta^4 + \gamma^4\,z^4}\right)} \, \bar{\Phi}_{2n}\left(\tilde{f}(z)\right) \qquad (28)$$

$$\bar{\Psi}_{2n+1}(z) = \sqrt{\left(\frac{\alpha\,z}{\beta^4 + \gamma^4\,z^4}\right)} \, \bar{\Phi}_{2n+1}\left(\tilde{f}(z)\right).$$

## 9. Trigonometric PTP Case

Lastly, it is possible to construct the final configuration of the target system concerning the uniform energy spectrum of the bound states as

$$\tilde{U}(z) = -\left(\frac{U_{01}}{\tilde{U}_0}\right)z^{-4} + U_{02}\left(1 + \tilde{U}_0\,z^4\right)$$

$$-\left[\left(\frac{21}{8\,\alpha^2}\right)z^4 + \frac{5/(32\,\sigma)}{z^4} + \frac{5/(32\,\sigma)}{\beta^4\,\gamma^4}\right]$$

$$\tilde{\Psi}_n(z) = \sqrt{\left(\frac{\alpha\,z}{\beta^4 + \gamma^4\,z^4}\right)} \, \tilde{\Phi}_n\left(\tilde{f}(z)\right). \qquad (29)$$

## 10. PCT Calculations For Mass Distribution – III

In this section, a new position-dependent mass function is proposed as

$$m = m_3(z) = \left(\frac{\alpha z^2}{\beta^2 + \gamma^2 z^6}\right)^2, \tag{30}$$

with the mass parameters $\alpha$, $\beta$, and $\gamma$. Thus its map function takes the form as

$$y = \hat{f}(z) = \int \sqrt{m_3}\, dz$$

$$= \frac{\alpha}{3\beta\gamma} \arctan\left(\frac{z^3}{3\beta\gamma}\right), \tag{31}$$

where the mapping function is $z = \left[(3\beta\gamma)\tan\left(\frac{3\beta\gamma}{\alpha} y\right)\right]^{1/3}$.

## 11. STP Case

In this case, the target potential and the wave fuctions for the symmetric and anti-symmetric cases for the uniform energy spectra in (14) and (15) can be obtained in compact form

$$U(z) = \hat{U}_0\, z^6$$

$$-\frac{1}{\alpha^2}\left[\frac{4}{z^6}(\beta^2 + \gamma^2 z^6)^2 - 3\gamma^2 (\beta^2 + \gamma^2 z^6) + 9\gamma^4 z^6\right]$$

$$\Psi_{2n}(z) = \sqrt{\left[\frac{\alpha z}{(\beta^2 + \gamma^2 z^6)}\right]}\, \Phi_{2n}\left(\hat{f}(z)\right) \tag{32}$$

$$\Psi_{2n+1}(z) = \sqrt{\left[\frac{\alpha z}{(\beta^2 + \gamma^2 z^6)}\right]}\, \Phi_{2n+1}\left(\hat{f}(z)\right),$$

where $\hat{U}_0 = (\kappa\, U_0)$ with $\kappa = \left(\frac{81\, \beta^4\, \gamma^4}{\alpha^2}\right)$.

## 12. SCP Case

By means of the PCT procedure, resulting system will become

$$\bar{U}(z) = \hat{U}_0\, z^{-6}$$

$$-\frac{1}{\alpha^2}\left[\frac{4}{z^6}(\beta^2 + \gamma^2 z^6)^2 - 3\gamma^2(\beta^2 + \gamma^2 z^6) + 9\gamma^4 z^6\right]$$

$$\bar{\Psi}_{2n}(z) = \sqrt{\left[\frac{\alpha z}{(\beta^2 + \gamma^2 z^6)}\right]} \, \bar{\Phi}_{2n}\left(\hat{f}(z)\right) \tag{33}$$

$$\bar{\Psi}_{2n+1}(z) = \sqrt{\left[\frac{\alpha z}{(\beta^2 + \gamma^2 z^6)}\right]} \, \bar{\Phi}_{2n+1}\left(\hat{f}(z)\right).$$

It is pointed out here that the spectrum of this system and that of its original counterpart are identical.

**13. Trigonometric PTP Case**

The mapping function gives rise to the following configuration having the same eigenspectra

$$\widetilde{U}(z) = \left[-\frac{U_{01}}{\widetilde{U}_\omega} z^{-6} + U_{02}\big(1 + \widetilde{U}_\omega z^6\big)\right]$$

$$-\frac{1}{\alpha^2}\left[\frac{4}{z^6}(\beta^2 + \gamma^2 z^6)^2 - 3\gamma^2(\beta^2 + \gamma^2 z^6) + 9\gamma^4 z^6\right]$$

$$\widetilde{\Psi}_n(z) = \sqrt{\left[\frac{\alpha z}{(\beta^2 + \gamma^2 z^6)}\right]} \, \widetilde{\Phi}_n\left(\hat{f}(z)\right), \tag{34}$$

where the parameter is $\widetilde{U}_\omega = (\omega \, U_0)$ with $\omega = \left(\frac{\alpha^2}{81 \, \beta^4 \, \gamma^4}\right)$.

**Concluding Remarks**

    This research article mainly deals with explicit and analytical solutions of the position-dependent effective mass (PDEM) Schrödinger equation for some quadratic form of trigonometric type potentials such as *tangent, cotangent* and *Pöschl-Teller* cases by implementing the mathematical procedure known as the point canonical transform (PCT). It also exhibits target (constructed) potentials for the original counterpart with mass distribution functions having some particular forms. Moreover, carrying out the PCT calculations allows us to establish both the energy spectra of the bound states and the corresponding eigenfunctions as well as canonical form of the current potentials algebraically. It is possible to here note that these results can also be adapted for description of several quantum few- and many-body physical systems including position-dependent unstable masses. Furthermore, they are applicable for analyzing some problems or issues encountered particularly in the fields of atomic and molecular, solid-states and condensed matter physics as well as examining the various devices or mechanisms with spatially changing of the mass configurations described in quantum optics and physical chemistry.

    Use of the PCT method enables us to express the mapping function. It plays a distinctive role for determining the target system, which includes compact forms of eigenspectra and wavefunctions. It is important to note that the energy levels of the reference (source) potential and the resulting ones have similar properties. Especially, the former and the target equivalent ones share the same energy eigenvalues. However, it can be noticed that former eigenfunctions and their counterparts (target potentials) have different forms. As a final remark, both the forms of the original and the spatial

dependence properties of the mass distribution functions $m = m(z)$ ensure the ultimate decision for the system whether it is fully solvable or not solvable.

**Acknowledgement**

The author would like to thank everybody who has contributed to him throughout progressing his academic career, especially his *mother*, *father* as well as his all lovely *family members*.